\documentclass[twocolumn]{rps-esrel2023}

\def\papername{\jobname}

\usepackage{natbib}

\begin{document}

\markboth{Julitz, T. M.; Schlüter, N. and Löwer, M.}{Scenario-based Failure Analysis of Product Systems and their Environment}

\twocolumn[

\title{Scenario-based Failure Analysis of Product Systems and their Environment}

\author{Tim Maurice Julitz}
\address{Department of Product Safety and Quality Engineering, University of Wuppertal, Germany. \email{julitz@uni-wuppertal.de}}

\author{Nadine Schlüter}
\address{Department of Product Safety and Quality Engineering, University of Wuppertal, Germany. \email{schlueter@uni-wuppertal.de}}

\author{Manuel Löwer}
\address{Department of Product Safety and Quality Engineering, University of Wuppertal, Germany. \email{loewer@uni-wuppertal.de}}

\begin{abstract} 
During the usage phase, a technical product system is in permanent interaction with its environment. This interaction can lead to failures that significantly endanger the safety of the user and negatively affect the quality and reliability of the product. Conventional methods of failure analysis focus on the technical product system. The interaction of the product with its environment in the usage phase is not sufficiently considered, resulting in undetected potential failures of the product that lead to complaints. For this purpose, a methodology for failure identification is developed, which is continuously improved through product usage scenarios. The use cases are modelled according to a systems engineering approach with four views. The linking of the product system, physical effects, events and environmental factors enable the analysis of fault chains. These four parameters are subject to great complexity and must be systematically analysed using databases and expert knowledge. The scenarios are continuously updated by field data and complaints. The new approach can identify potential failures in a more systematic and holistic way. Complaints provide direct input on the scenarios. Unknown, previously unrecognized events can be systematically identified through continuous improvement. The complexity of the relationship between the product system and its environmental factors can thus be adequately taken into account in product development. 
\end{abstract}


]

\section{Introduction}\label{sec1}

\subsection{Background and motivation}
Timely identification of potential failures in the early phases of product development poses enormous challenges for companies. As the development process progresses, the cost of correcting a fault becomes more expensive the longer it takes to discover it. \cite{Boehm1984}. A technical product system is in permanent interaction with its environment during the usage phase. \cite{Hitchins.2007}. This interaction can lead to failures that significantly endanger the safety of the user and negatively affect the quality and reliability of the product. This was illustrated by the well-known Tesla car accident, in which the automated vehicle control system could not distinguish between the bright sky background and the white side of the truck, resulting in the death of the Tesla driver. \cite{NationalTransportationSafetyBoard.2017}. The early investigation of failures that occur during interactions between products and their environment is therefore of high research relevance and can have fatal consequences in extreme cases. With the expansion of functional safety standards in 2022, scenario analyzes are now also standard in the automotive sector to convert unknown events into known events. \cite{iso21448}.

\subsection{Research problem and objective}
Various failure analysis methods (e.g. FMEA or FTA) are used in product development to prevent failures. However, these methods do not adequately or specifically address the product-environment interaction that occurs during product usage, resulting in undetected potential failures of the product which lead to complaints. A scenario method focused on the usage phase can depict this interaction. \cite{Kurtoglu.2008}, \cite{Amer.2013}. So far, the focus has been on strategic product planning, \cite{Villamil.2022}; requirements engineering, \cite{Liu.2012}, \cite{Scarinci.2019} or design of product services. \cite{Geum.2011}. As a consequence, the development of a failure identification concept through continuously improved product usage scenarios is required. A model-based use of scenarios is able to identify potential failures that can arise from the product-environment interaction at an early stage of product development. The continuous improvement of scenarios is important to systematically identify unknown events in the usage phase. This can be done using feedback from complaints and field data, which have not been sufficiently considered in current scenario-based methods.

\subsection{Research questions and hypothesis}
This study aims to develop a failure identification concept that addresses the product-environment interaction during the usage phase through the continuous improvement of product usage scenarios. To achieve this goal, the following research questions are addressed:

\begin{itemlist}
\item{R1.}  How can the product-environment interaction during the usage phase be effectively addressed in failure analysis methods?
\item{R2.}  How can scenario-based methods be improved in order to systematically identify potential failures that can arise from product-environment interaction at an early stage of product development?
\item{R3.}  How can actual failures from the usage phase be integrated into the continuous improvement of scenarios for failure identification in the product-environment interaction?
\end{itemlist}

This is based on the hypothesis that a model-based use of scenarios that systematically takes into account the product-environment interaction during the usage phase and includes actual failure information from complaints and field data, leads to more effective and holistic failure identification in product development and leads to a reduction in complaints. Section \ref{sec2} of the paper gives a brief overview of the state of the art of scenario-based failure analysis. Sec. \ref{sec3} introduces the approach of the new method. In sec. \ref{sec4} the integration of tools for the application takes place. Sec.\ref{sec5} and \ref{sec6} discuss the results and give an outlook and a conclusion.

\section{Scenario-based failure analysis in the literature}\label{sec2}

\subsection{Scenario failure analysis}
\cite{Arogundade.2020} shows that scenarios are not widely used in failure analysis. So far, it is better known as a strategic planning method that can be used to explore possible future situations and development paths. \cite{Dean.2019}. \cite{Amer.2013} emphasizes that scenarios are suitable for assessing uncertainties in the case of complexity.

Some approaches and research results can be found in the literature on scenario-based failure analysis. A selection is presented below. For example, there are research papers on a scenario-based FMEAs by \cite{Kmenta.2000} or \cite{Issad.2017}. These scenarios are limited to the technical product system, since only the interactions within the component and functional level are analyzed. The causal and logical relationships between failures were presented in a fault tree by linking events that lead to failure scenarios. \cite{Tekinerdogan.2008}. Event Tree Analysis (ETA) is well suited for scenario-based failure analysis. Each ETA is also a scenario analysis as the possibility of different events is assessed from a starting point. \cite{Zoric.2022}. A combination of FMEA and QFD (Quality Function Deployment) focuses on data-driven development considering failure modes with digital twin use cases. \cite{Newrzella.2022}. 

\subsection{Product-environment interaction in scenarios}
The failure analysis approaches mentioned above do not take the environment into account or only insufficiently or not systematically. In \cite{Maier.2018}, the product system environment is modeled using a MBSE use case approach for a VR environment. The views  ``human'', ``product'' and ``enviornment'' are linked via structure and behaviour models. A database is created for each property, which can be expanded manually. Failures are only detected here by observing the VR scene. \cite{Zhang.2020} takes a similar approach, dividing the product system into the ``physical'', ``human'', ``cyberspace'' and ``external environment''. From this, various possibilities of interaction are modeled. It is emphasized that design knowledge and experience are important. \cite{Schuh.2014} describes the environmental factors through various indicators that take quantitative values to assess their probabilities. However, a failure analysis is not carried out with these approaches. The product-environment interaction can also be described by a cube with four squares. \cite{Bielefeld.2020}. The system is modeled and linked by the product system, environmental factors, effects and events. Each of the four elements is created from a database and expert knowledge. Failure analysis can be implemented by reducing the complexity to only those system elements that result in a functional failure. For example, a fault tree analysis can now be carried out for the identified elements. However, the question remains which features need to be considered and whether the databases are complete and reflect all possible environmental interactions? To ensure this, information from the usage phase must be fed back in order to continuously improve the scenarios. This does not happen with this approach. 

\subsection{Improvement of scenario models}
Continuous improvement of scenario models is important to identify events in the product usage phase that are unknown during product development. The original scenario management approach from \cite{Gausemeier.1998} provides a scenario development step after the scenario forecast. In this case, updating the scenarios involves identifying inconsistencies in the projection, which are then removed from the original model. \cite{Villamil.2022} proposes a cyclical approach, with expert workshops and feedback sessions scheduled in each cycle and a (1) backcasting, (2) forecasting, and
(3) bridging process in the scenario development. In \cite{Maier.2018}, the model is adjusted after observing the created VR scene. Only Gausemeier's approach is systematic in this regard. However, a connection between failures and environmental factors is not discussed here.

\section{Method Development Approach} \label{sec3}
\subsection{Methodology Overview}
\begin{figure}
\centering
  \includegraphics[width=\linewidth]{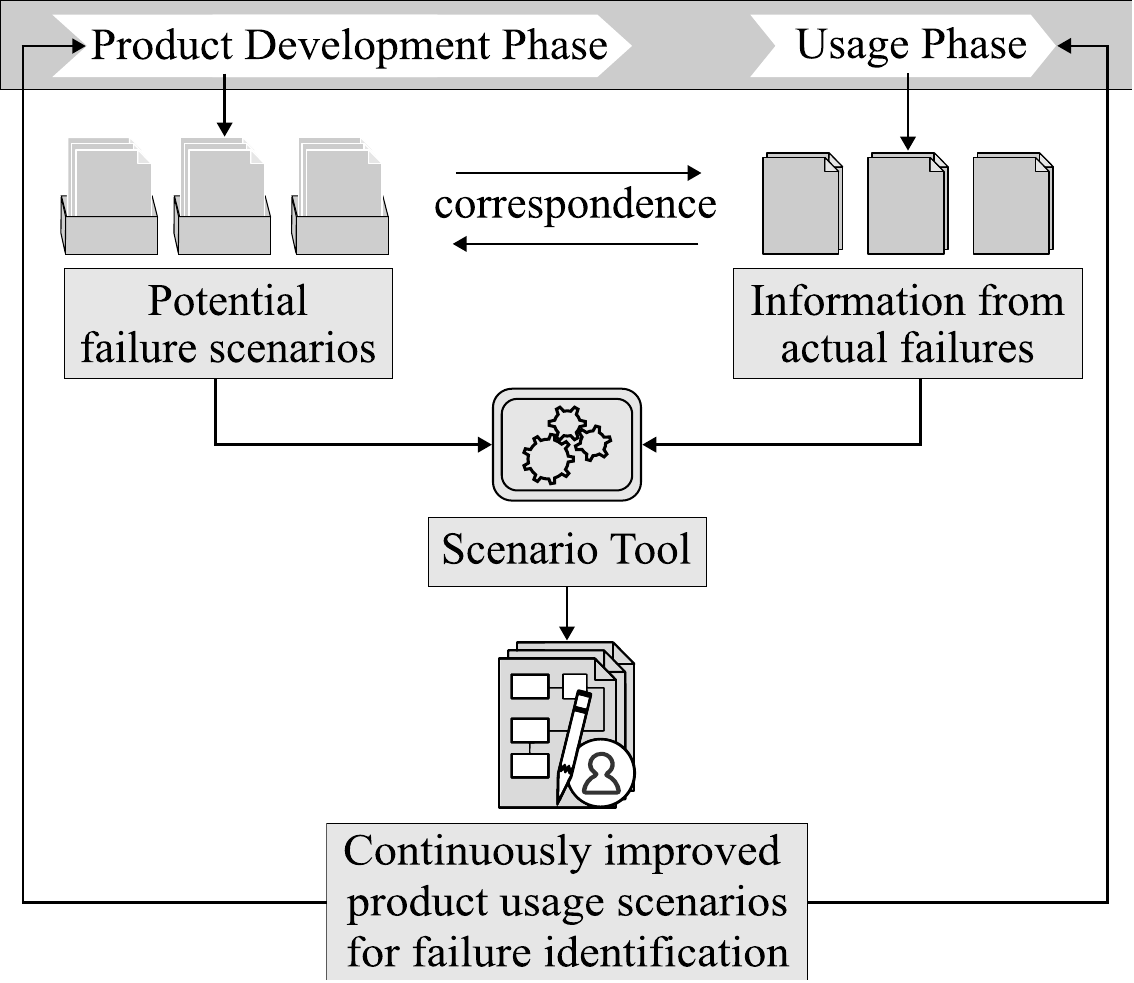}
  \caption{Failure identification approach through continuously improved product usage scenarios.}
  \label{fig:method}
\end{figure}

A method for failure identification through continuously improved product usage scenarios is developed~(Fig.~\ref{fig:method}). It enables the analysis of potential failures resulting from the interaction of the product system and its environment in the use phase and provides continuously updated information for product development.

The scenario-based failure analysis from product development is compared with real failure information from the product usage phase. The level of correspondence between potential and actual failure data is the basis for continuous improvement of the scenario tool. The potential failures are determined based on the approach of \cite{Bielefeld.2020}. Actual failure information from product usage are collected after~\cite{Ansari.2020}. The output of both approaches need to be unified to determine the degree of agreement between potential and actual failures~(sec.~\ref{fusla}). The updated information are evaluated in the scenario tool in order to be able to evaluate new, previously unknown scenarios~(sec.~\ref{tool}). The two approaches were chosen because they use a similar system modeling approach, which makes them excellent to combine with each other, and because they adequately map the interaction of the product system environment in a model-based approach with database support.

\subsection{Product Development Scenarios and Actual Failures from Usage}
In order to be able to make a statement about the effectiveness of failure identification in the early phases of product development, it must first be worked out how the degree of correspondence between scenarios from product development and actual failures from the usage phase can be determined. Complaints can be used for this purpose. \cite{Ansari.2020}. Given the variety of scenarios and complaints, this is a particular challenge. Not only that their individual characteristics are very different, but also that the product development and the customer have a different view and a different understanding of the scenarios and complaints. Accordingly, it seems necessary to identify the individual features and mark possible interfaces. To achieve the sub-goal, various existing methods are adopted and linked synergistically.

\subsection{Failure Analysis Scenario Tool}
Once the degree of correspondence between scenarios and complaints has been recorded, potential failures in the interaction between the product system and the environment must be systematically identified in the early phases of the product development. A scenario-based expert tool is to be developed for this purpose. For the development of the tool, the findings from the usage phase (complaints) should be used to create scenarios and update them continuously.

\subsection{Continuous Improvement of Product Usage Scenarios}
After the respective scenarios have been revised, the next step is the necessary adaptation and optimization of the failure identification method for failure analysis in product development with a focus on potential failures that result from the interaction of product systems and environmental influences in the usage phase. The adjustments must be made according to a continuous improvement process so that the failure analysis and failure identification scenarios can be constantly improved. This reduces uncertainties in the product development failure analysis, which leads to an increase in the product quality of technical product systems and ensures user and customer safety.

\section{Refinement of the Method} \label{sec4}
\subsection{Product-Environment System Model in the Scenario Tool} \label{tool}
\begin{figure}
  \centering
  \includegraphics[width=.9\linewidth]{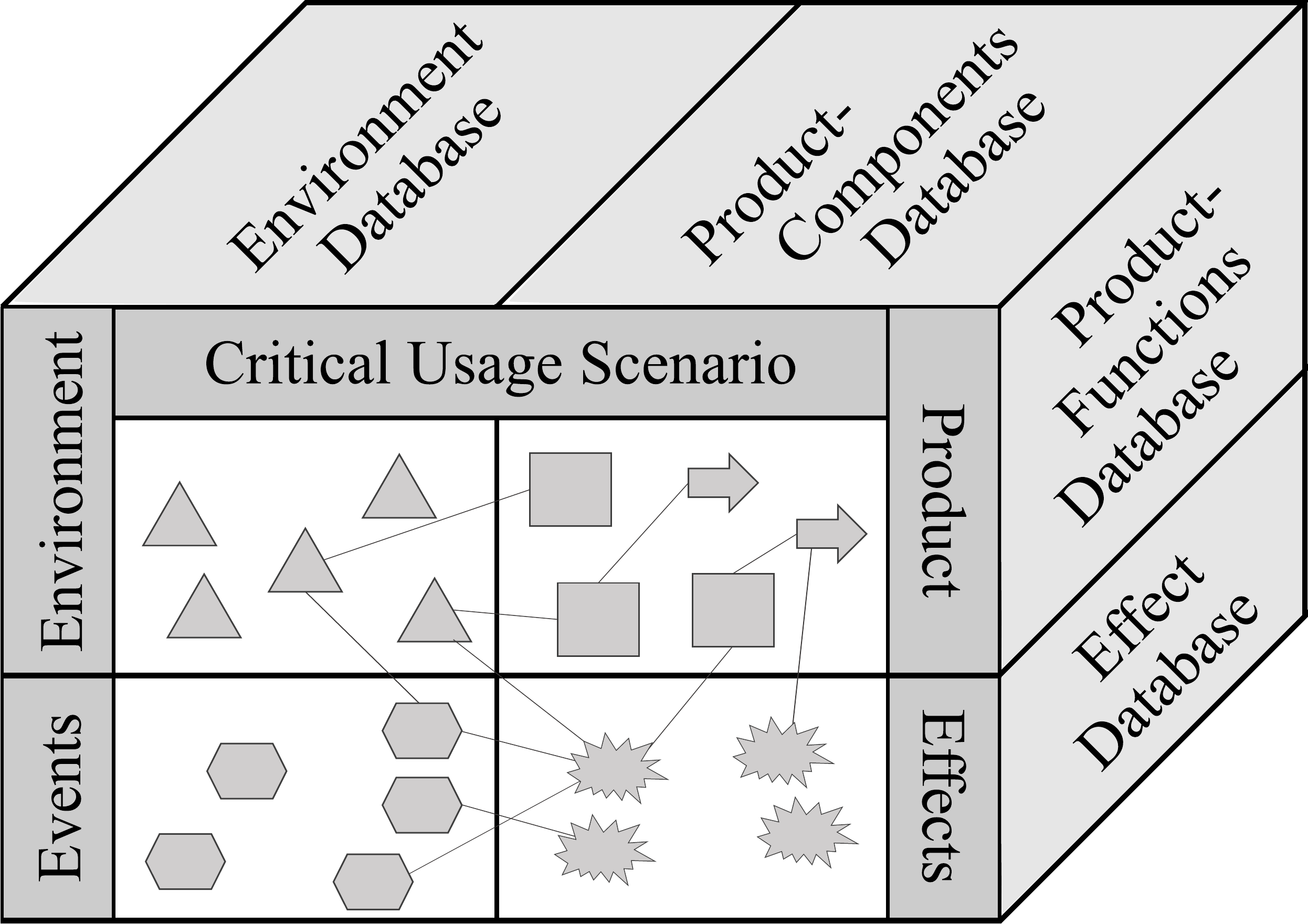}
  \caption{Surface of the Scenario Tool according to \cite{Bielefeld.2020}}
  \label{fig:cube}
\end{figure}

Connections between different influencing factors as well as combinations of failures and their consequences can be analyzed using a system model of the product. \cite{Haberfellner.2019}. Consequently, Fig.~\ref{fig:cube} represents the tool as a model of the product system, with a focus on the functions and system components as proposed in \cite{DIN.EN.60812}. This can build on \cite{Bielefeld.2020} work on holistic failure analysis. 

In order to describe the system holistically, the views ``product'', ``environment'', ``events'' and ``effects'' are modeled and linked. Information on each view is stored and in constantly updated databases. They contain all eventualities that can occur in the product-environment system. 

Effects represent the connection between a failure cause and a failure consequence. Effects are e.g. vibration or thermal energy. They can be logical, physical or mathematical. The temporal sequence of events is part of the definition of scenarios. Unexpected events (e.g. car accidents, explosions, etc.) can have a significant negative impact on the application process. In order to limit complexity, the tool surface only concentrates the information of individual critical usage processes. This contains isolated functions of the product including the associated system elements.

\subsection{Formalized failure network analysis} \label{kausal}
\begin{figure}
\centering
  \includegraphics[width=\linewidth]{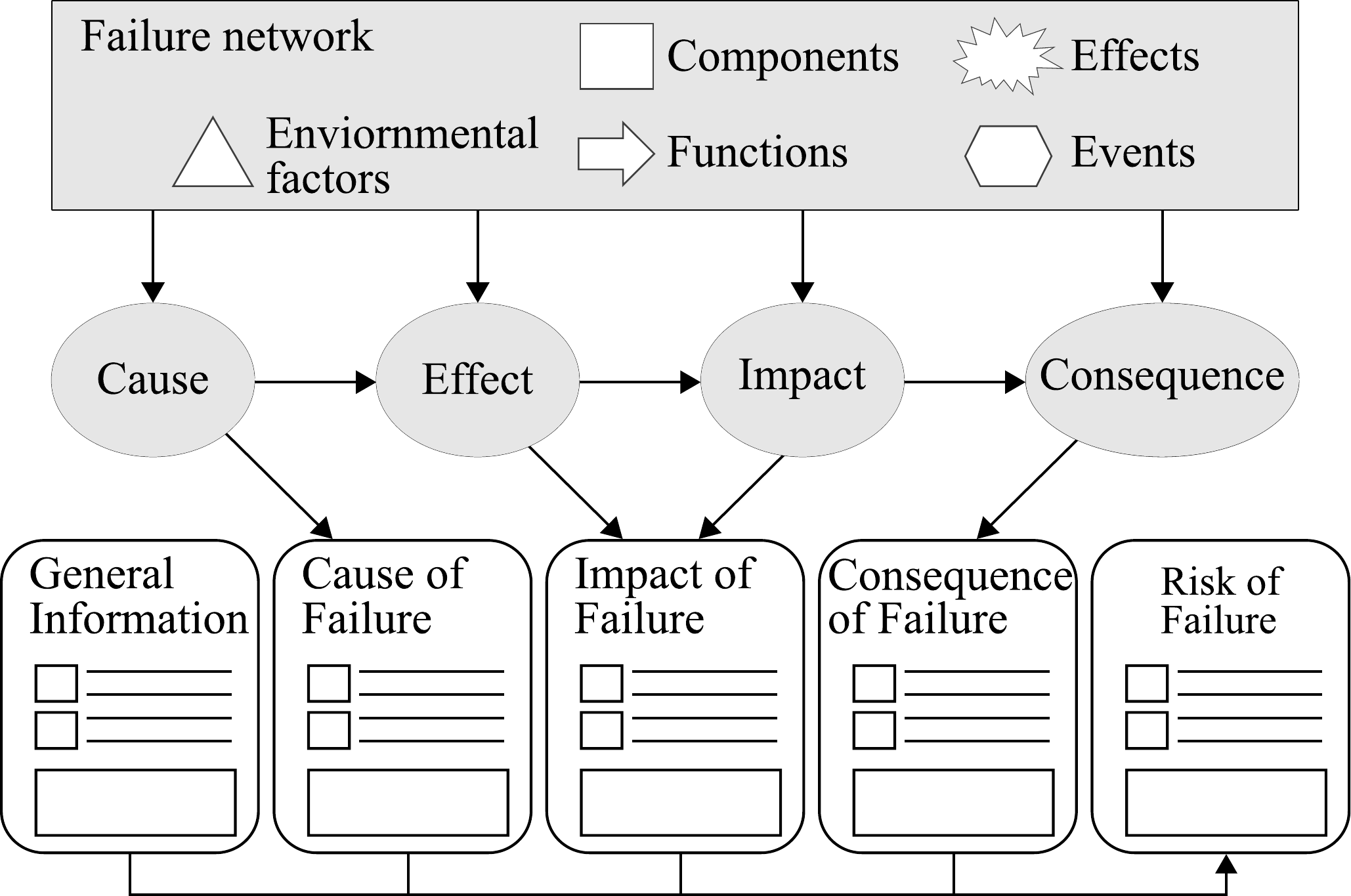}
  \caption{Formalized failure network analysis based on \cite{Bielefeld.2020}}
  \label{fig:chain}
\end{figure}

\cite{Bielefeld.2020} emphasized that functions have major importance when describing a failure, while the element ``effect'' connects both failure and function. In this case, the failure definition refers to the inability to perform a necessary function. It can be analyzed by examining the cause of the failure, the effect, its impact, and the resulting consequences (see Fig.~\ref{fig:chain}). This sequence can help to identify the root cause of the failure, prevent future occurrences, and minimize the negative impact and outcomes.

The failure network is modelled according to the elements in Fig.~\ref{fig:cube}. Once a critical usage scenario is identified, the linked system elements can be analyzed. Several functions can also be assigned to one use case. Bielefeld (2020) created a VBA script that facilitates the transfer of information from system elements to the failure chain. This script also extracts additional information by querying the system elements (components, functions, events, effects, environmental factors) and stakeholders. This process leads to a formalized failure description and an assessment of the associated risk. The failure types are identified and assigned to the system elements. This allows, for example, the association of a cascading component failure with a malfunction resulting from a specific scenario.  Experts can use this information to make improvements to the product. This step is carried out in the product development phase where there is limited information about the performance of the product in the usage phase. Scenarios in which the product misbehaves may not be detected.

\subsection{Validation of potential failures by a complaint generalization algorithm} \label{fusla}
\begin{figure}
\centering
  \includegraphics[width=.85\linewidth]{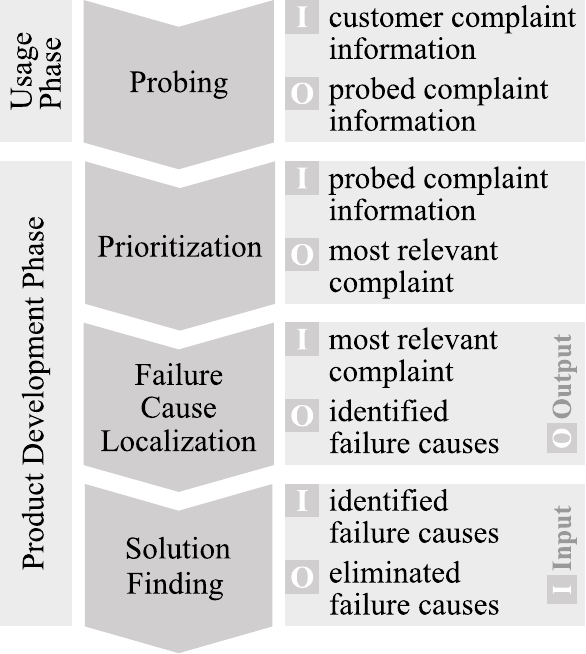}
  \caption{Failure-Cause-Searching and Solution Finding Algorithm (Fusla). \cite{Heinrichsmeyer.2019}
}
  \label{fig:fusla}
\end{figure}

To eliminate this uncertainty, the databases of the possible specifications of the product usage scenario elements (Fig.~\ref{fig:cube}) must be constantly updated with information from the usage phase. For this purpose, an algorithm for generalizing customer complaints with integrated failure cause and solution finding can be used (Fig.~\ref{fig:fusla}). 

If failure scenarios occur that were not recognized during product development, the scenario tool databases are updated so that the new scenario can be taken into account from now on. Unknown scenarios become known step by step. The starting point of the algorithm are complaint texts written by the customer. \cite{Ansari.2020}. A generalization and probing is done by the tokenization technique from the field of Natural Language Processing (NLP). The priority is calculated quantitatively, followed by a failure-cause localization, which consists of mapping the failure description into the elements of the product system (here: requirements, components, functions, processes and people) and suggesting corrective actions based on failure cause categories. \cite{Heinrichsmeyer.2019}.

In order to validate potential failures from the failure network analysis with actual failures from the failure-cause search and solution-finding algorithm, the output of both approaches must be unified. The possibility of linking consists in the description of the system elements with the subsequent classification into failure cause categories. In addition to the system elements in Fig.~\ref{fig:cube}, the algorithm (Fig.~\ref{fig:fusla}) also uses the elements ``actors", ``processes" and ``requirements". The failure information can be classified into the following categories with their refinement in parentheses: General Information (description), Cause of failure (function, component, requirement, process, environment, event), impact of failure (function, component, requirement, process, environment, event, effect), consequence of failure (technical product, human, stakeholder), Risk (calculated value). The attributes of these categories are stored in continuously updated databases. The use of databases is explicitly important for events, effects and environmental factors, since the number of these elements and their interaction can reach a very high level of complexity. The risk can be calculated for both tools according to the VBA script from \cite{Bielefeld.2020}. Based on the unified information, the degree of correspondence is calculated.

\subsection{Workflow assembly}
\begin{figure}
\centering
  \includegraphics[width=\linewidth]{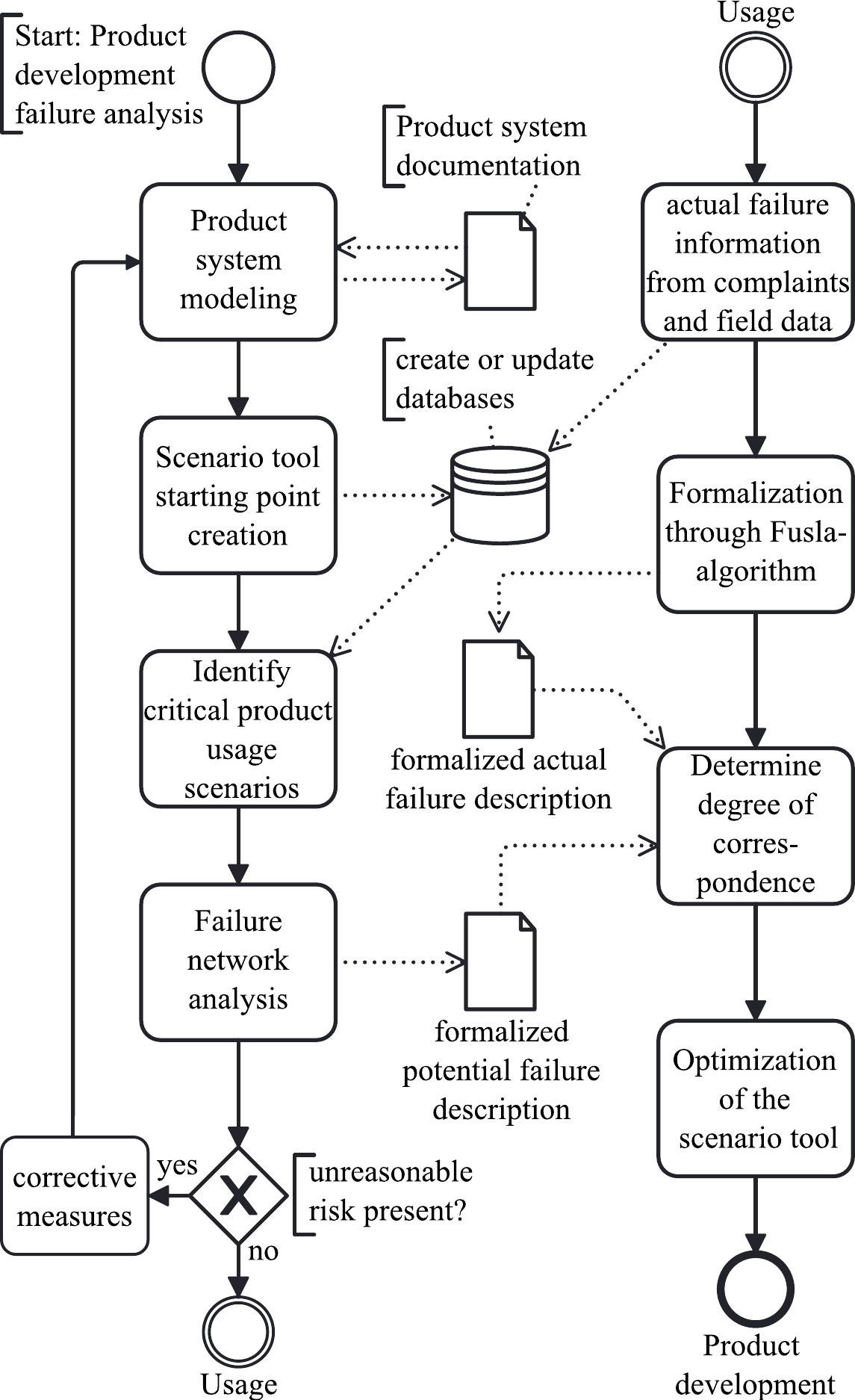}
  \caption{Workflow of the method for scenario-based failure identification of product systems and their environment
}
  \label{fig:bpmn}
\end{figure}

The workflow of the method for scenario-based failure identification of product systems and their environment is summarized in Fig.~\ref{fig:bpmn}.

The activities in the product development phase are shown on the left. The usage phase is addressed on the right-hand side. The first three steps on the left side are part of section~\ref{tool}. A starting point is needed when creating the scenario from scratch. This means that the databases are filled from the documentation for the first time and the connections between the elements are drawn. The ``Failure network analysis" refers to section~\ref{kausal}. The right side is described in section~\ref{fusla}. Both phases, product development and usage are linked by continuously updated documents and databases, that serve to identify, prevent and correct failures.

\section{Discussion} \label{sec5}
The failure analysis in product development could be connected to actual failures from the usage phase in order to continuously improve the developed method. Scenarios were used to holistically analyze the technical product system and its environment. For this purpose, a tool was used that models critical usage scenarios by linking the views of product, environment, effects and events. Potential failure networks of these views are analyzed to get a formalized failure descriptions. Continuous improvement is achieved by comparing this potential formalized failure description to actual formalized failure descriptions generated using an NLP-based algorithm for handling customer complaints. The deviation is determined using the ``degree of correspondence" indicator. If new, previously unrecognized failure scenarios are identified, the scenario tool databases are updated so that this information can be used in product development. However, the method has the disadvantage that an improvement effect only occurs after a few cycles, since the databases are initially empty. In order to ensure holistic failure identification, the method would need an initialization phase. A solution might be to use accessible databases like the triz effects database or internal databases as a starting point. This would be difficult to implement, especially for potential events. Therefore, methods are required that make it possible to fill the databases with holistic and systematic scenario descriptions right from the start. The databases are currently filled with static models such as structure diagrams. For a more holistic scenario identification, the implementation of dynamic models like state machines or activity diagrams could be considered. SysML could be used to connect dynamic and static models. Furthermore, a software implementation of the tool has to be implemented in which the user receives an output through a defined input. Finally, the method should be validated using application examples. The effectiveness of the method can be measured using the indicator ``degree of correspondence".

\section{Conclusion} \label{sec6}
A method has been developed that can identify potential failures at an early stage of product development. It could be shown that the product-environment interaction can be addressed by using scenario techniques in failure analysis by a model based approach (R1). Existing scenario-based methods were adapted and further developed to achieve this goal (R2). By continuously improving the scenario tool with formalized failure information, potential failures in product development are identified more systematically and holistically (R3). This reduces the risk of not recognizing failures that may result in harm or damage during the usage of the product. The part of the hypothesis that the method leads to a reduction in complaints has not yet been confirmed, as a test phase is necessary. It was found that the method gradually leads to an improvement in failure identification, but still has uncertainties in the beginning. Especially for this initial phase, methods still have to be developed that can systematically describe the large number of possible failure scenarios. Since this work mainly focused on scenario-based failure analysis using failure networks, other methods of failure analysis should be considered. So far, the failure networks have been combined with fault tree analysis. A combination with FMEA, event trees, bow-tie analyzes or fish-bone diagrams is conceivable. A key challenge is the initial filling of the databases of the scenario tool, which will be addressed in future research. In addition to static models, dynamic models such as state machines or activity diagrams should also be used.

\begin{acknowledgement}
This research is funded by the Deutsche Forschungsgemeinschaft (DFG, German Research Foundation) – Project No. 502859764.



\bibliographystyle{chicago}
\bibliography{References}
\end{acknowledgement}

\end{document}